\documentstyle[dina4,doublespace,11pt]{article}

\def\r{\ref}
\def\t{\gamma}
\def\P{\Psi}
\def\l{w}
\def\baa{\begin{array}}
\def\eaa{\end{array}}
\def\ba{\begin{eqnarray}}
\def\ea{\end{eqnarray}}

\def\be{\begin{equation}}
\def\ee{\end{equation}}
\def\la{\label}
\def\c{\cite}

\def\x{\xi}
\def\xb{\bar{\xi}}
\def\f{\frac}

\def\bi{\bibitem}
\def\Hx{H^{(\x)}}
\def\Hxb{H^{(\xb)}}
\def\H{{\cal H}}
\def\p{\partial}
\def\0{|p \rangle }
\def\k{2k}
\def\ka{\kappa}

\begin{document}
\begin{center}
\LARGE{An Integrable Model of Quantum Gravity}
\vskip2cm
\large{D.~Korotkin$^{\star\dagger}$\footnote{Supported
by Alexander von Humboldt Foundation.}  and
H.~Nicolai$^{\star}$}

\vskip0.5cm
$^{\star}$
II. Institute for Theoretical Physics, Hamburg University, \\
Luruper Chaussee 149, 22761 Hamburg, Germany \\
$^{\dagger}$
Steklov Mathematical Institute, Fontanka, 27, St.Petersburg, 191011,
Russia
\end{center}
\vskip2.0cm
{\bf ABSTRACT.} We present a new quantization scheme for $2D$ gravity
coupled
to an $SU(2)$ principal chiral field and a dilaton; this model
represents
a slightly simplified version of stationary axisymmetric quantum gravity.
The analysis makes use of
the separation of variables found in our previous work \c{1} and is
based on a two-time Hamiltonian approach. The quantum constraints
are shown to reduce to a pair of compatible first order equations,
with the dilaton playing the role of a ``clock-field".
Exact solutions of the Wheeler-DeWitt equation are constructed via the
integral formula for solutions of the
Knizhnik-Zamolodchikov equations.

\vskip1.0cm
{\bf Introduction.}
At present only few models of quantum gravity exist for which the
Wheeler-DeWitt equation can be solved exactly.
The known examples include pure gravity \c{Witten} and
supergravity \c{dWMN} in three dimensions, as well as certain
mini-superspace models such as static spherically symmetric
gravity \c{KTh} or locally supersymmetric Bianchi-type
models \c{Graham}. All of these models describe only
finitely many physical degrees of freedom and are thus
bound to miss essential features of quantum gravity.
It is therefore clearly desirable to find models of quantum gravity with
infinitely many physical degrees of freedom, which can still be
treated exactly. In this paper we present such a model describing
$2D$ gravity coupled to a principal chiral $SU(2)$ model and
a dilaton, and demonstrate
that it can be quantized exactly by use of methods
developed in the context of $2D$ integrable systems \c{Korep, Faddeev}.
The construction is based on our previous work \c{1}, where we have
proposed a novel canonical formulation of the Ernst equation based
on the complete separation of the equations of motion
in the isomonodromic sector of the theory
and the use of the logarithmic derivative of the related
$\Psi$-function with respect to the spectral parameter
as the fundamental canonical variable.
Our model can be regarded as a version of the
``midi-superspace" approximation of quantum gravity, but
differs from dimensionally reduced gravity in that we have replaced the
non-compact configuration space $SU(1,1)/U(1)$
by the compact group manifold $SU(2)$.
Apart from this technical modification (the reasons for which will
be briefly explained at the end of this paper)
the model captures all the essential
features of dimensionally reduced quantum gravity, and is a non-trivial
example of a completely integrable model of quantum gravity with
infinitely many propagating degrees of freedom.
A further intriguing aspect of the present work is the
possible relevance of quantum groups suggested by the link with
the Knizhnik-Zamolodchikov (KZ) equations.

{\bf Classical model.}
For definiteness we consider only Euclidean worldsheets
with complex local coordinates $(\x,\xb)$ (the extension of our results
to Lorentzian worldsheets is completely straightforward).
The worldsheet metric is
\be
ds^2=h(\x,\xb)d\x d\xb \;\;\;\;\;\;\;\;\;\;h\equiv e^{\k}
\la{m}\ee
and the Lagrangian is taken to be
\begin{equation}
{\cal L} =\rho \Big( R + {\rm tr} (g_{\x} g^{-1}
g_{\xb}g^{-1}) \Big)
\la{L}\end{equation}
Here $\rho(\x,\xb)\in {\bf R}$ is the dilaton field originating from
dimensional reduction as explained e.g. in \c{Nic},
and the chiral field
$g(\x,\xb)\in G=SU(2)$ describes the matter content of the model
(subscripts stand for partial derivatives).
The conformal factor in (\r{m}) represents a further,
but dependent, degree of freedom and appears in (\r{L})
via the Gaussian curvature of the worldsheet; in the conformal gauge
this term becomes
$ R=(\log h)_{\x\xb}/h. $
The equation of motion for $\rho(\x,\xb)$ following from (\r{L}) reads
\be
\rho_{\x\xb}=0
\la{dil}\ee
and possesses the general solution
$\rho(\x,\xb) = f(\x)+f(\xb) $.
Assuming $f\neq const$
we can perform an analytic local change
of coordinates such that $f(\x)=\x/2i$ (global aspects of this
change of variables were discussed in \c{2}, but will not concern us here).
In the sequel we shall thus always assume
\be
\rho(\x,\xb)={\rm Im}\, \x
\la{gaugefix}\ee
and also set $x(\x,\xb ):={\rm Re} \, \x$.
We will take some liberty in referring to $\rho$ and
$x$ as the ``time" and ``space" coordinates, respectively. The gauge choice
(\r{gaugefix}) is analogous to the light cone gauge in string theory
whereby one identifies the longitudinal target space coordinate $X^+$
with the world sheet time coordinate.
The remaining (and compatible) equations of motion read
\be
\k_{\x}=\f{\x -\xb}{4}{\rm tr}(g_{\x}g^{-1})^2\;\;\;\;\;\;\;\;
\k_{\xb}=\f{\xb - \x}{4}{\rm tr}(g_{\xb}g^{-1})^2
\la{h}\ee
and
\be
 \big( (\x -\xb ) g_{\x}g^{-1}\big)_{\xb}
+ \big( (\x -\xb ) g_{\xb}g^{-1} \big)_{\x} = 0.
\la{ee}\ee
As is evident from these equations, the $2D$
gravitational fields and the matter fields are coupled through the
dilaton $\rho$; furthermore it can be shown that the obvious
solution $\rho= const$ of
(\r{dil}) would imply $g_{\x}=g_{\xb}=h_{\x}= h_{\xb}=0$,
i.e. the trivial solution for the matter fields, in which case the
gravitational sector would become purely topological (strictly speaking,
this argument relies on the positive definiteness of the Cartan Killing
metric on the group $G$). Therefore the model possesses no
non-trivial flat space limit since the matter fields
act as sources for $2D$ gravity and thus distort the
two-dimensional background geometry.

Equation (\r{ee}) is the compatibility condition of the following linear
system \c{M}:
\be
\f{\p\Psi}{\p\x}+\f{\t}{\x-\xb}\f{1+\t}{1-\t}\f{\p\Psi}{\p\t}
=\f{g_{\x}g^{-1}}{1-\t}\P\;\;\;\;\;\;\;\;
\f{\p\Psi}{\p\xb}+\f{\t}{\xb-\x}\f{1-\t}{1+\t}\f{\p\Psi}{\p\t}
=\f{g_{\xb}g^{-1}}{1+\t}\P
\la{UV}
\ee
where $\P \equiv \P(\x,\xb ;\t )$ is a two-by-two matrix,
from which the solution of (\r{ee}) can be reconstructed, and
$\t$ is a ``variable spectral parameter"
\be
\t (\x , \xb ; w) =
\f{2}{\x-\xb}\bigg\{\l-\f{\x+\xb}{2}+\sqrt{(\l-\x)(\l-\xb)}\bigg\}
\la{gamma}    \ee
with $\l\in {\bf C}$ a constant of integration which may be regarded
as the ``hidden" spectral parameter. Notice that the poles in the
$\t$ plane appearing on the LHS of (\r{UV}) are produced by differentiation
of $\t$ with respect to $\x$ and $\xb$, respectively.

According to the approach developed in \c{1}, the fundamental quantities
are the logarithmic derivatives of $\Psi$ with respect to the
spectral parameter $\t$:
\be
\Psi_{\t}= A(\t)\Psi\;\;\;\;\;\;\;\;\;\;
A(\t)\equiv \sum_{j}\f{A_j}{\t-\t_j}
\la{Pt}
\ee
where for notational clarity we have suppressed the dependence
on the coordinates $(\x , \xb )$.
The two-by-two matrices $A_j(\x,\xb)$ are the residues at
$\t_j=\t(\x,\xb ;\l_j)$ in the complex $\t$-plane, with constants $\l_j$.
We here consider only the isomonodromic sector, for which the
sum (\r{Pt}) is finite; in general, however, the sum (\r{Pt}) may contain
an infinite or even continuous number of terms (if the sum is replaced
by an integral).

The compatibility conditions of (\r{Pt}) and (\r{UV}) yield the
following
system of differential equations
for the matrices $A_j(\x,\xb)$ \c{1}:
\be
A_{j\x}= \f{2}{\x -\xb} \sum_{k\neq j} \f{[A_k,A_j]}{(1-\t_k)(1-\t_j)}
\;\;\;\;\;\;\;\;
A_{j\xb}=\f{2}{\xb- \x} \sum_{k\neq j} \f{[A_k,A_j]}{(1+\t_k)(1+\t_j)}
\la{1}
\ee
As emphasized in \c{1} these equations are automatically
compatible unlike the original linear system (\r{UV}).
The problem of solving the nonlinear partial differential equation
(\r{ee}) is thus reduced to solving a system
of {\it ordinary} matrix differential equations.
Once the solutions $A_j$ of (\r{1}) are known, the chiral field
$g(\x,\xb)$
can be recovered from $A(\gamma )$  by specializing the
spectral parameter $\t$ to the values $\gamma = \pm 1$, viz.
\be
(\x -\xb ) g_{\x} g^{-1}=  2A(\t ) \big|_{\gamma =1} \;\;\;\;\;\;\;
(\x -\xb ) g_{\xb} g^{-1}=  2A(\t )\big|_{\gamma =-1} ;
\la{g}
\ee
again we have not displayed the
dependence on the coordinates in these expressions.
The conformal factor $h(\x,\xb)$ may then be computed by
integration of (\r{h}). In \c{1} we have shown that it is essentially
the $\tau$-function associated with (7) in the sense of \c{Jimbo}.

All results so far are actually valid for arbitrary $SL(2,{\bf C})$
(and, in fact, $GL(n,{\bf C})$) valued
matrices $g(\x,\xb)$, but we will consider only $G=SU(2)$ in the remainder
of this paper. The solutions of the system (\r{1}) corresponding
to $g\in SU(2)$ and $h\in {\bf R}$ are characterized by the following

\newtheorem{Theorem}{Lemma}
\begin{Theorem}
Let $\big\{ A_j , t_j ; j=1,...,N\big\}$ be invariant with respect to the
involution $A_j\rightarrow -A_j^{\dagger}\;,\;\; \t_j\rightarrow
-\bar{\t}_j$ (i.e. if $\t_j=-\bar{\t}_j$ then $A_j\in su(2)$; if
$\t_j=-\bar{\t}_k \;,\;k\neq j$ then $A_j=-A_k^{\dagger}$). Then
the constants of
integration in (\r{h}) and (\r{g}) may be chosen in such a way
that $g\in SU(2)$ and $h\in {\bf R}$.
\end{Theorem}

The conditions of the Theorem 1 in particular imply
that $A(1)^\dagger = A(-1)$. In the sequel we restrict attention
to the case $\t_j=-\bar{\t_j}$
(i.e. $w_j\in {\bf R}$), so that $A_j\in su(2)$ for all $j$.
Regularity of $\Psi (\t)$ at $\t=\infty$ together with
(\r{Pt}) implies a further constraint, viz.
\be
\sum_{j=1}^{N} A_j =0
\la{constr}\ee
This condition is related to the asymptotical flatness of
the corresponding classical solutions.
All constraints are consistent with the equations of motion (\r{1}) and
form a closed algebra with respect to the Poisson bracket (\r{PB}) below.

It was also explained in \c{1} that the equations (\r{1}) can be
interpreted as
a two-time hamiltonian system with respect to ``times"
$\x$ and $\xb$. The relevant Poisson brackets are given by
\be
\{A(\t)\;\stackrel{\otimes}{,}
A(\mu)\}=\Big[ r(\t-\mu), A(\t)\otimes I + I\otimes A(\mu)\Big]
\la{PB}
\ee
where the classical $r$-matrix $r(\t)$ is equal to $\Pi/\t$ with $\Pi$ the
permutation operator in ${\bf C}^2\times {\bf C}^2$. The dynamics in
the $\x$ and $\xb$-directions is governed by the complex Hamiltonians
\be
\Hx= \f{1}{\x - \xb}{\rm tr}\,\Big( A^2(1) \Big) \;\;\;\;\;
{\rm and} \;\;\;\;\;
\Hxb= \f{1}{\xb- \x}{\rm tr}\, \Big(  A^2(-1) \Big)
\la{H12}
\ee
respectively, which depend explicitly on $(x,\rho )$
through the prefactor $\rho^{-1}$ and the spectral values $\t_j$
via (\r{gamma}). Observe that
we have $(\Hx )^\dagger = \Hxb$ by Theorem 1.
As expected from the
decoupling of variables in (\r{1}), the flows generated by $\Hx$ and
and $\Hxb$ commute, i.e. $\{ \Hx , \Hxb \} =0$, and this ensures that
the equations of motion can be consistently integrated.
Remarkably, the dimension of the system has been effectively reduced
by trading the ``space" variable $x$ and the
``time" variable $\rho$ for two ``time" variables $\x$ and $\xb$.
We can thus neglect the dependence on
$x$ and $\rho$ for all practical purposes and regard the spectral parameter
currents $A(\gamma )$ at a fixed but arbitrarily chosen
base point $(x_0, \rho_0 )$ as the fundamental canonical variables.

In view of (\r{H12}) we can reexpress (\r{h}) in the form
\be
\k_{\x} - \Hx = 0 \;\;\;\;\; , \;\;\;\;\;
\k_{\xb} - \Hxb = 0      \la{WDW}
\ee
Below, we will have to interpret these equations
as canonical constraints \`a la Dirac rather than merely as equations
determining the conformal factor; upon quantization, they will
yield the Wheeler-DeWitt equation and the diffeomorphism constraint.

{\bf Quantization.}
To quantize the model, we replace the Poisson
brackets (\r{PB}) by commutators in the usual fashion:
\be
[A(\t)\stackrel{\otimes}{,}A(\mu)]
= i\hbar [r(\t -\mu)\;,\;A(\t)\otimes I +
I\otimes A(\mu)]
\la{CR}\ee
The matrix elements of $A(\t )$ thus become operators acting on a
Hilbert space to be specified below;
note that on the LHS of (\r{CR}),
we have a commutator of operators in Hilbert space
whereas on the RHS we have a commutator of ordinary matrices.
This means in particular that the expansion (\r{Pt}) is no longer valid
as an operator statement, but must be reinterpreted as a property of
the states on which $A(\t )$ acts.
The expressions for the Hamiltonians
(\r{H12}) remain unchanged. We write
\be
A(\t)\equiv \f{i\hbar}{2}
        \left(\baa{cc}  h(\t)\;\;\;\;\;\;\;\;\;\;\; 2e(\t)\\
                        2f(\t)\;\;\;\;\;\;\;\; - h(\t)\eaa\right)
\ee
As is well known the Hamiltonians (\r{H12}) can be diagonalized by means
of the Bethe ansatz method \c{Skl}. For this purpose, one considers a
lowest weight state $\0$ labeled by some analytic
function $p = p(\t )$ subject to $\overline{{p(\t)}} = - p(-\bar \t )$.
This state is assumed to obey
$h(\t) \0 = p(\t) \0 $ and $f(\t) \0 =0$. The Bethe states are then
\be
|p;v_1,...,v_M \rangle := e(v_1)...e(v_M) \0  \la{Bethe}
\ee
The parameters $v_1,...,v_M$ must satisfy certain constraints called
Bethe equations ($M$ is the number of excitations) in order for
the Hamiltonians (\r{H12}) to act diagonally on them; however, we
will not need these constraint equations below.
The off-diagonal operators $e(\t)$ and $f(\t)$
play the role of creation and annihilation operators,
respectively (interchanging the role of $e(\t)$ and $f(\t)$ turns the
representation ``upside down"). The classical expansion
(\r{Pt}) corresponds to the special choice
\be
p (\t) = \sum_{j=1}^N \f{ m_j}{\t - \t_j}
\la{at}
\ee
By fixing the number of poles in this fashion,
we have restricted the full Hilbert space to its $N$ soliton sector,
which can be represented as a direct product
${\H}^{(N)} = \H_1 \otimes \dots \otimes \H_N$.
Only on this subspace can we expand the operator
$A(\t)$ in analogy with (\r{Pt}), with
\be
A_j\equiv  \f{i\hbar}{2}
\left(\begin{array}{cc}  h_j\;\;\;\;\;\;\;2e_j\\
2f_j\;\;\;\;\;\; - h_j\end{array}\right)
\la{spin1}
\ee
where $h_j\;,e_j,\;f_j$ are
the standard Chevalley generators of $SU(2)$ obeying
the commutation relations
$$
[h_j,\;e_j]= 2e_j\;\;\;\;\;\;\;[h_j,\;f_j]= -2f_j\;\;\;\;\;\;
[e_j,\;f_j]= h_j
$$
as a consequence of (\r{CR}). By well known arguments, unitarity
then requires each $\H_j$ to be a representation space of $SU(2)$
with highest weight $m_j \in {\bf Z}$.
We may therefore view the quantized
$N$ soliton sector as a system of $N$ spins located at the points
$(x,\rho) =(w_j,0)$ where the classical solutions of the system
(\r{h}), (\r{ee}) have singularities on the worldsheet \c{K}.
Since the full Hilbert space is the direct sum of its $N$ soliton
sectors, its structure resembles that of
a Fock space with $N$ playing the role of a particle number operator
(not to be confused with the number $M$ of excitations in(\r{Bethe})).

The central task is now to solve the quantum analog of the
constraints (\r{WDW}), which are nothing but linear combinations of
the Wheeler-DeWitt equation and the diffeomorphism constraint
for our model (corresponding to translations in $\rho$ and $x$,
respectively), and read
\be
\Big( \k_{\x} -  \Hx \Big) \Phi = \Big( \k_{\xb} - \Hxb \Big) \Phi = 0.
\la{WDW1}
\ee
where $\Phi$ is the full quantum state
(so-called ``wave function of the universe"). Accordingly, we
extend the phase space by the
gravitational degrees of freedom $\k_{\x}\;,\;\k_{\xb}$ and
the  variables $\x$, $\xb$.
To proceed we must first construct an operator representation
for these new phase space variables. From the
canonical brackets given in \c{1} we deduce the commutators
$$
[ \k_{\x} , \x ] = [\k_{\xb} , \xb ] = i\hbar  \;\;\; , \;\;\;
[ \k_{\x} , \xb ] = [\k_{\xb} , \x] = 0
$$
We take
$$
\k_{\x} = i\hbar \f{\p}{\p \x} \;\;\; , \;\;\;
\k_{\xb} = i\hbar \f{\p}{\p \xb} \la{cofa}
$$
The main advantage of this choice is that by
representing $(\x, \xb )$ as multiplication operators
we salvage their interpretation as coordinates (otherwise the spectral
parameter $\t$ would not remain a function but become a non-local
differential operator and thus very awkward to deal with).
It is then obvious that the two equations (\r{WDW1})
are mutually compatible for the same reason that their classical
counterparts (\r{h}) are. Recall that the worldsheet coordinates
$(\x , \xb )$ appear explicitly only because we have adopted the special
gauge (\r{gaugefix}) identifying the dilaton field with one of the
coordinates. In other words, this
choice of gauge makes the quantum state $\Phi$
time-dependent through the identification of time with
the ``clock field" $\rho$. We note that this long suspected mechanism
for the emergence of time from the ``timeless" Wheeler DeWitt equation
here comes almost for free (see e.g. \c{Isham} for a review and further
references). In a covariant treatment the gauge choice (\r{gaugefix})
would have to be undone,
and the full quantum state would be a functional
of $\rho$ rather than a function of the worldsheet coordinates.
The fact that through this choice of gauge the field
variables $\k_{\x}$ and $\k_{\xb}$, which originated from the conformal
factor, become canonically conjugate to the worldsheet coordinates,
is also in accord with the interpretation of $\k_{\x}$ and
$\k_{\xb}$ as longitudinal target space momenta suggested in \c{Nic,2},
and can be construed as evidence for a ``stringy" interpretation
of the theory, with our treatment being analogous to the quantization
of the string in the lightcone gauge.

The equations (\r{WDW1}) now take the form
\be
i\hbar\f{\p\Phi}{\p\x}= H^{(\x)} \Phi\;\;\;\;\;\;\;
i\hbar\f{\p\Phi}{\p\xb}= H^{(\xb)} \Phi
\la{WDW2} \ee
where in the $N$-soliton sector $\Phi$ is an ${\H}^{(N)}$-valued
function of $(\x , \xb )$.
The explicit solutions of (\r{WDW2}) may be obtained by exploiting
a surprising link between
(\r{WDW2}) and the KZ equations \c{KZ}
determining the $N$-point correlation functions of the WZNW model,
which read
\be
\f{\p\tilde \Phi}{\p\t_j}=\f{1}{\ka}\sum_{k\neq j}\f{\Omega_{jk}}
{\t_j-\t_k}\tilde \Phi
\la{KZ}\ee
with an ${\H}^{(N)}$-valued function $\tilde \Phi$, where
$\Omega_{jk}=\f{1}{2} h_j\otimes h_k + f_j\otimes e_k +
e_j\otimes f_k\; (j\neq k)\;$
is a linear operator acting non-trivially only in
$\H_j$ and $\H_k$.
In \c{SV} (see also \c{B,Resh1,Var}) it was shown that the
general solution of (\r{KZ}) can be obtained in terms of the following
integral representation over Bethe eigenstates
\be
\tilde \Phi =\int_{\Delta}
    W(\t,v) \; |p;v_1,...,v_M \rangle dv_1...dv_M
\la{Sol}\ee
where
$$W(\t,v)=\prod_{1\leq k < j\leq N} (\t_j-\t_k)^{m_j m_k/2\ka}
\prod_{1\leq s < r\leq M} (v_r-v_s)^{2/\ka}
\prod_{r=1}^{M}\prod_{j=1}^{N}(\t_j-v_r)^{-m_j/\ka}   $$
and $\Delta$ is a family of cycles in $v$-space which
do not intersect the hyperplane where $W$ becomes singular;
furthermore, $W$ should be single-valued on $\Delta$.
For the precise definition of $\Delta$
and the proof of (\r{Sol}) see \c{SV,B,Resh1,Resh}.
It is important that the parameters $v_r$ in (\r{Sol}) are {\em not}
subject to the Bethe equations.
In \c{B} it was shown that the state (\r{Sol}) is annihilated by
the constraint (\r{constr}) iff
\be
M=\f{1}{2}\sum_{j=1}^{N} m_j
\la{M}\ee
Using (\r{gamma}) and (\r{spin1})
we can now verify that the KZ equations (\r{KZ}) imply that
\be
\Phi = \prod_{j=1}^{N} \bigg( \f{\partial \t_j}{\partial w_j}
      \bigg)^{i\hbar \f{m_j (m_j+1)}{4}} \tilde \Phi  \la{Sol1}
\ee
satisfies (\r{WDW2}), provided the state
$\Phi$ has vanishing ``total spin" in accordance with (\r{constr})
and we make the identification
\be
\ka = -1/i\hbar .   \la{kappa}
\ee

{\bf Outlook.}
To conclude we would like to briefly
expose the extra complications that one faces when
attempting to quantize dimensionally reduced gravity, whose
distinct feature is the {\em non-compactness} of the configuration space
inhabited by the propagating degrees of freedom. As we
already mentioned, the relevant equations of motion can be formally
obtained from our equations
(\r{h}), (\r{ee}) by replacing the compact $SU(2)$ manifold
by the non-compact coset space $SU(1,1)/U(1)$, after which
(\r{ee}) becomes just the Ernst equation. Due to the
division by the maximal compact subgroup, which is already
necessary at the classical level in order to
prevent the energy from being unbounded below, we now encounter
new canonical constraints necessary to ensure the decoupling
of unphysical negative norm states. In the case at hand, it turns out
that, in fact, one must impose {\em infinitely many} such constraints
corresponding to the maximal compact subgroup of the Geroch group
$\widehat{ SU(1,1)}$. Moreover,
the representation spaces for each ``spin"
will be infinite-dimensional (unlike the spaces $\H_j$ above) because of
the non-compactness of the group $SU(1,1)$.

\end{document}